\documentclass[prb,aps, twocolumn,floatfix,showpacs,showkeys, superscriptaddress]{revtex4-1}

\usepackage{graphicx}
\usepackage{epstopdf}
\usepackage{amssymb}
\usepackage{color}

\begin{document}

\title{Crossover from charge density wave stabilized antiferromagnetism to superconductivity in Nd$_{1-x}$La$_x$NiC$_2$ compounds}

\author{Marta Roman}
\email{marta.roman@pg.edu.pl}
\affiliation{Faculty of Applied Physics and Mathematics, Gdansk University of Technology,
Narutowicza 11/12, 80-233 Gdansk, Poland}

\author{Leszek Litzbarski}
\affiliation{Faculty of Applied Physics and Mathematics, Gdansk University of Technology,
Narutowicza 11/12, 80-233 Gdansk, Poland}

\author{Tomasz Klimczuk}
\affiliation{Faculty of Applied Physics and Mathematics, Gdansk University of Technology,
Narutowicza 11/12, 80-233 Gdansk, Poland}
 
\author{Kamil K. Kolincio}
\email{kamkolin@pg.edu.pl}
\affiliation{Faculty of Applied Physics and Mathematics, Gdansk University of Technology,
Narutowicza 11/12, 80-233 Gdansk, Poland}
\affiliation{RIKEN Center for Emergent Matter Science (CEMS), Wako, Saitama 351-0198, Japan}

\begin{abstract}

The path from the charge density wave antiferromagnet NdNiC$_2$ to a noncentrosymmetric superconductor LaNiC$_2$ has been studied by gradual replacement of Nd by La ions. The evolution of physical properties has been explored by structural, magnetic, transport, magnetoresistance and specific heat measurements. 
With the substitution of La for Nd, the Peierls temperature is gradually suppressed, which falls within the BCS mean-field relation for chemical pressure with a critical concentration of $x_c$ = 0.38. As long as charge density wave is maintained, the antiferromagnetic ground state remains robust against doping and despite of a Néel temperature reduction shows a rapid and sharp magnetic transition. Once the CDW is completely suppressed, intermediate compounds of the Nd$_{1-x}$La$_x$NiC$_2$ series reveal symptoms of a gradual softening of the features associated with AFM transition and increase of the spin disorder. Immediately after the antiferromagnetic transition is depressed to zero temperature, the further incorporation of La ions results in the emergence of superconductivity. This crossover in the Nd$_{1-x}$La$_x$NiC$_2$ is discussed in the terms of the possible quantum critical point.
\end{abstract}

\maketitle

\section{Introduction}
\indent
The family of the ternary rare-earth dicarbides RNiC$_2$ (R - rare-earth metal) crystallizing in the noncentrosymmetric orthorombic CeNiC$_2$-type crystal structure ($Amm2$)\cite{Jeitschko1986} has recently been extensively studied due to the variety of ground states which they offer. This family is known to exhibit, depending on the rare-earth (R) atom, the  charge density wave (CDW) at Peierls temperatures $T_{CDW}$ ranging from 89 K for PrNiC$_2$\cite{Yamamoto2013} to around 450 K for LuNiC$_2$\cite{Roman2018_1, Steiner_2018}, superconductivity (SC), ferromagnetism (FM) or antiferromagnetism (AFM) at low temperatures. So far the CDW state, which in RNiC$_2$ compounds is associated with the Ni atom chains constituing a quasi-low dimensional electronic structure, has been found in most RNiC$_2$ members (R = Pr-Lu)\cite{Murase2004, Laverock2009, Wolfel2010, Sato2010, Ahmad2015, Michor2017, Roman2018_1}. Recent studies revealed the linear scaling of the Peierls temperature with unit-cell volume for R = Sm-Lu \citep{Roman2018_1}. The magnetism in RNiC$_2$, however originates entirely from the rare-earth sublattice through the Ruderman-Kittel-Kasuya-Yosida (RKKY) interaction between local magnetic moments mediated by conducting electrons associated with the Ni atoms carrying no magnetic moment themselves\cite{Schafer1997, Kotsanidis1989}. With the exception R = (Y, La, Pr, Sm, Lu), all the RNiC$_2$ undergo an antiferromagnetic transition with Néel temperatures $T_N<$ 25 K \cite{Onodera1995, Onodera1998, Kotsanidis1989, Bhattacharyya2014, Pecharsky1998, Yakinthos1990, Hanasaki2011, Uchida1995, Matsuo1996}. Only a weak magnetic anomaly was observed for PrNiC$_2$\cite{Onodera1998, Kolincio2017}, while SmNiC$_2$ \cite{Onodera1998} and LaNiC$_2$ \cite{Wiendlocha2016, lee_superconductivity_1996, Pecharsky1998, Quintanilla2010, Landaeta2017} exhibit ferromagnetic and superconducting ground states, respectively. YNiC$_2$ and LuNiC$_2$ remain paramagnets above $T$ = 1.9 K \cite{Kotsanidis1989, Steiner_2018}. 
\\
\indent
The vast diversity of physical properties offered by the RNiC$_2$ family makes them a promising platform to explore interrelationships between different types of ordering, expecially between CDW, magnetism and superconductivity. The recently explored interplay between CDW and magnetism has been found to exhibit a bilateral character. On the one hand, the antiferromagnetic state has been suggested to be created, or at least substantially reinforced by the preexisting charge density wave state \cite{Hanasaki2017, Roman2018_2}. On the other hand, the same AFM state (NdNiC$_2$ and GdNiC$_2$) partially supresses the CDW\cite{Yamamoto2013, Lei2017, Kolincio20163, Kolincio2017} although it allows the coexistence of both entities. Moreover, a completely destructive influence of ferromagnetism on the CDW was observed in SmNiC$_2$ \cite{Shimomura2009, Hanasaki2012, Lei2017, Kim2012}. 
In contrast, in PrNiC$_2$ the magnetic anomaly has been found to have a constructive impact on the nesting properties \cite{Yamamoto2013, Kolincio2017}. 
In such a group of materials, an even more fertile field allowing the exploration of these interactions opens up when two competing magnetic or electronic ground states tend towards zero temperature and the quantum fluctuations corresponding to them collide at a quantum critical point (QCP)\cite{Kopp2005, Scalapino2012, Friedemann2009, Wang2018, Jang2018, Wang20181}. A quantum critical point could be revealed and thus explored by tuning the ground state via nonthermal parameters such as pressure, composition or magnetic field. The effect of pressure can be studied equivalently by applying external force or via chemical alloying, causing a change in the lattice parameters (increase or decrease, depending on the difference in atom size). The emergence of a ferromagnetic quantum criticality was previously suggested in SmNiC$_2$ studied under pressure \cite{Woo2013}, SmNiC$_{2-x}$B$_x$ \cite{Morales2014}, and Sm$_{1-x}$La$_x$NiC$_2$ solid solution \cite{Prathiba2016, Lee2017}. So far, the aniferromagnetic QCP in this family have been revealed  under hydrostatic pressure in LaNiC$_2$ \cite{Landaeta2017} and CeNiC$_2$ \cite{Katano_2019}. Alas, no signatures of quantum criticality have been observed in their solid solutions\cite{Katano_2017_1, Katano_2017_2}.
\\
\indent
LaNiC$_2$ is an unconventional superconductor below $T_{SC}$ = 3 K with magnetic fluctuations assisted Cooper pairs creation\cite{Landaeta2017}. The proximity of AFM state seen in NdNiC$_2$ at $T_N$ = 17 K (preceded also by a Peierls transition at $T_{CDW}$ = 121 K) and this type of superconductivity in the phase diagram of RNiC$_2$ motivated us to use chemical alloying to explore the path between NdNiC$_2$ and LaNiC$_2$ from the vantage point of the evolution of the underlying ground states and the possible quantum criticality at AFM-SC crossover.
In this paper, by means of structural, transport, magnetic and heat capacity measurements we investigate the influence of La doping of NdNiC$_2$ on charge density wave, antiferromagnetism and superconductivity. A comprehensive $T-x$ phase diagram showing putative AFM QCP near $x^*$ = 0.88 for Nd$_{1-x}$La$_x$NiC$_2$ ($0\leq x\leq 1$) series is constructed.

\begin{figure*}[ht]
\includegraphics[angle=0,width=2.1\columnwidth]{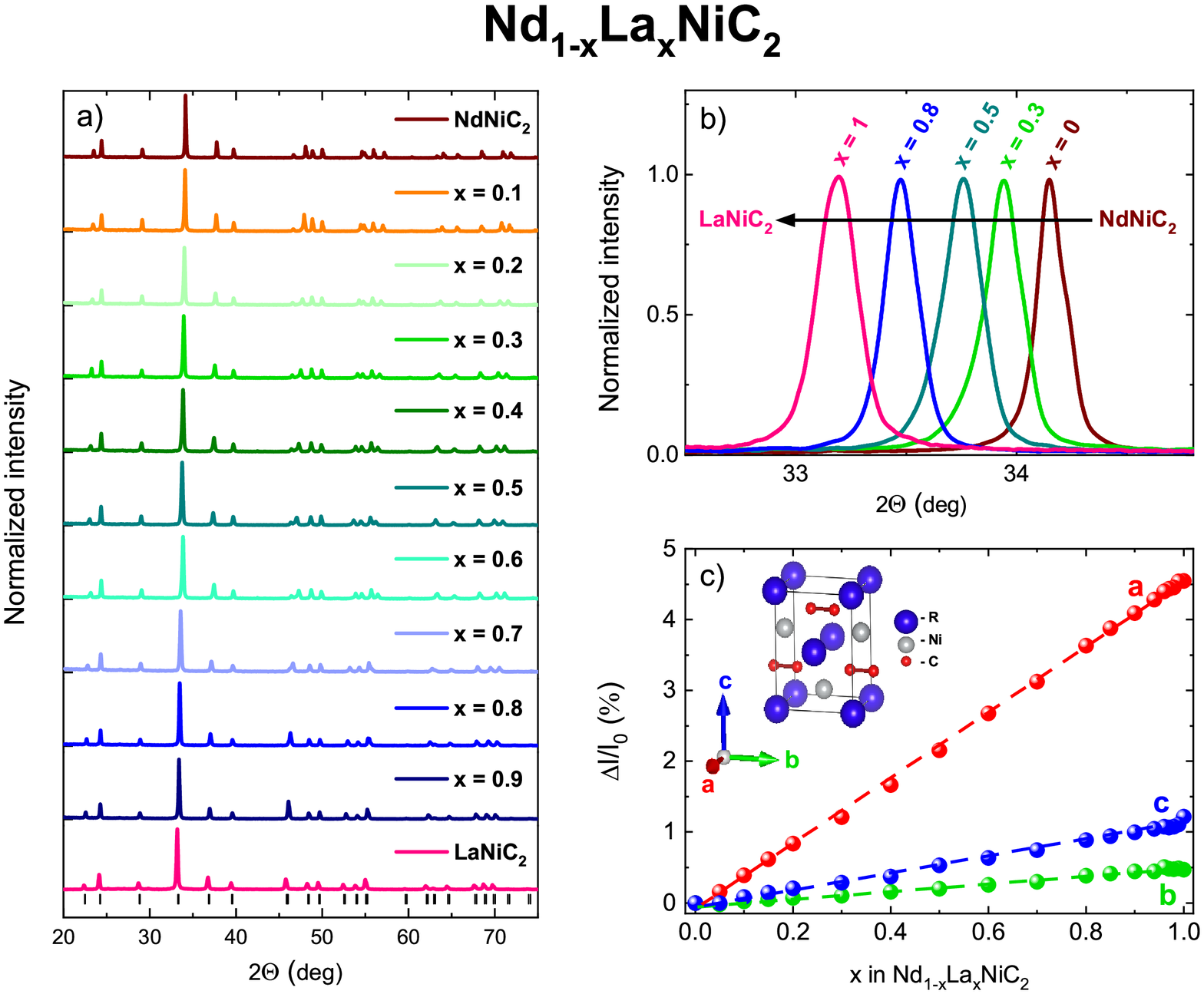}
 \caption{\label{XRD} a) Normalized pXRD patterns for selected samples from the Nd$_{1-x}$La$_x$NiC$_2$ series. Bragg peak positions for LaNiC$_2$ are marked with vertical ticks. b) Shift of the main (111) reflection with the change of $x$. c) Relative change of the lattice parameters $a$, $b$, and $c$ as a function of $x$.   }
  \end{figure*}

\section{Experimental}

\indent
The synthesis of the polycrystalline Nd$_{1-x}$La$_x$NiC$_2$ ($0\leq x\leq 1$)  series was performed via arc melting under a zirconium-gettered ultrapure argon atmosphere followed by further annealing at 900$^o$C for 12 days. The purity of the elements used was: Ni (3N), C (5N) and Nd (3N), La (4N); and due to the high volatility of the lanthanides and carbon, the 2\% of Nd and La, and 3\% of C excess was added in order to compensate for the loss during arc melting. The overall change of weight after the synthesis process was negligible ($\leq 1\%$) indicating that the elemental
concentration was close to the actual alloying level. The details of the whole procedure with the synthesis of other solid solutions were previously described in \cite{Roman2018_2}.  
\\
\indent
The phase purity of the samples from the whole series was confirmed by powder x-ray diffraction (pXRD) on a PANalytical X'Pert Pro diffractometer with a Cu K$_{\alpha}$ source. The lattice parameters were determined from a LeBail profile refinement of the diffraction patterns by using FULLPROF\cite{FULLPROF} software.

The transport properties, magnetic susceptibility and heat capacity were measured with Quantum Design Physical Properties Measurement System (PPMS) allowing the application of a magnetic field up to 9 T  in the temperature range from 1.9 to 300 K. Magnetization measurements were performed using the ac and dc magnetometry system (ACMS) option. The ac magnetization for superconducting samples was measured with a dc field of 5 Oe and 1 kHz excitations with a 3 Oe amplitude. The specific heat measurements were performed using a standard relaxation method. The electrical resistivity was measured with a regular four-probe technique with thin ($\phi=37 \mu m$) Pt wires playing the role of electric contacts, that were spark-welded to the polished surfaces of thin samples. The magnetoresistance  was measured with magnetic field applied perpendicularly to the current direction.

\section{Results and discussion}
Diffraction patterns of the powdered samples from the Nd$_{1-x}$La$_x$NiC$_2$ ($0\leq x\leq 1$) series were collected at room temperature and are depicted in Fig. \ref{XRD}.
All observed reflections are succesfully indexed in the orthorhombic CeNiC$_2$-type structure with space group $Amm2$ and no secondary phase was detected within the whole series. 
With increasing La content ($x$) in Nd$_{1-x}$La$_x$NiC$_2$ solid solutions, one can observe the shift of the Bragg reflection lines towards lower values of 2$\Theta$, which is consistent with replacing Nd$^{3+}$ ions with La$^{3+}$ having larger ionic radius (shift of the main (111) reflection is shown in Fig. \ref{XRD}b)).
The lattice parameters determined from the LeBail fit for the whole Nd$_{1-x}$La$_x$NiC$_2$ series and for parent compounds NdNiC$_2$ and LaNiC$_2$ are in good agreement with previous reports \cite{Roman2018_2, Prathiba2016}. As it is depicted in Fig.\ref{XRD}c), the unit cell parameters $a$, $b$ and $c$ follow a linear relationship with the La doping rate ($x$) and hence Vegard’s law is obeyed. The largest relative change is observed for $a$ parameter and reaches 4.5\% while $b$ change is barely noticeable (0.5\%) which is associated with the rigid bond between carbon dimers arranged along the $b$-axis (see the crystal structure picture in Fig. \ref{XRD}c))

 \begin{figure*}[ht]
\includegraphics[angle=0,width=2.1\columnwidth]{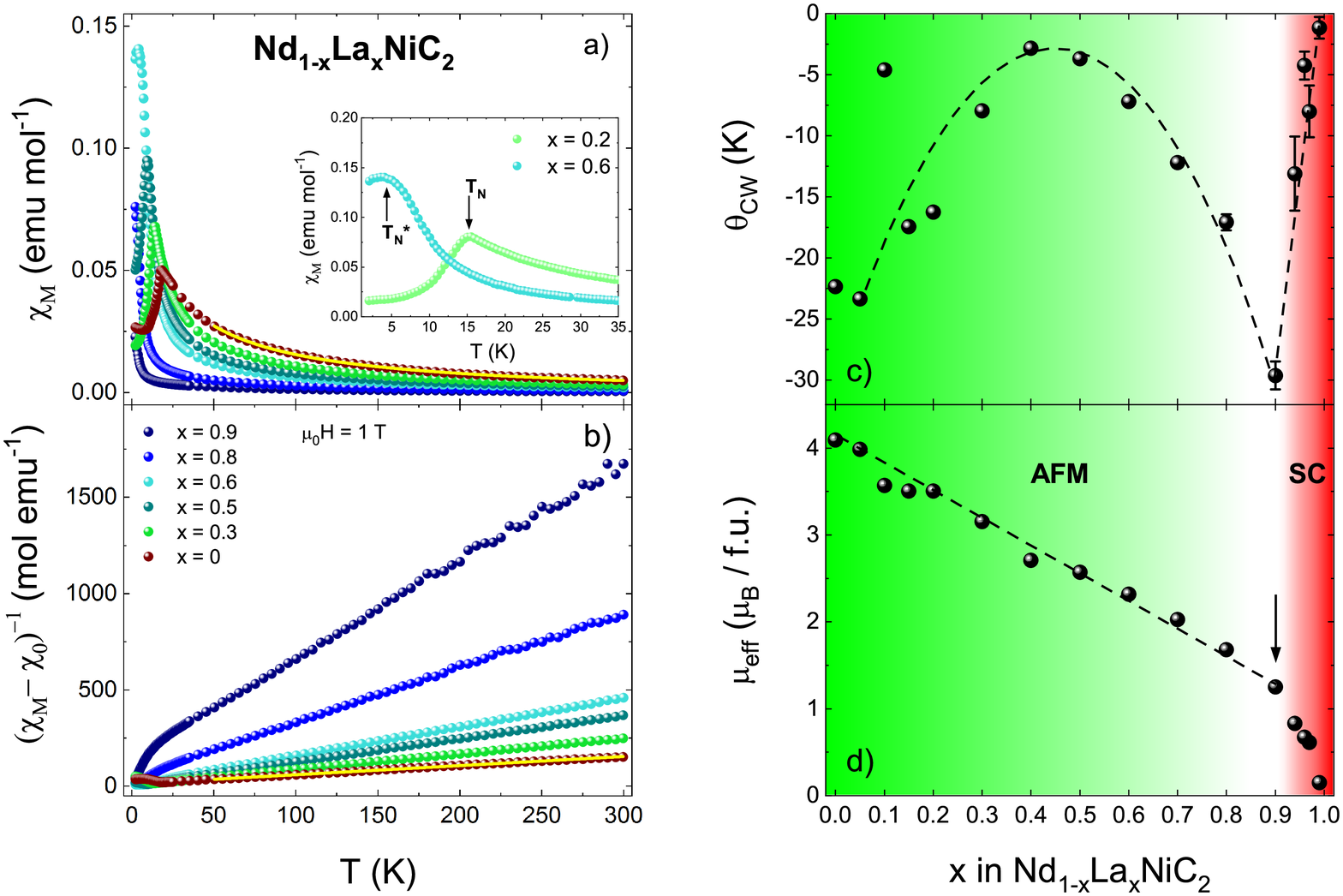}
 \caption{\label{AFM} a) Temperature dependence of the molar magnetic susceptibility $\chi_M(T)$ and b) of the reciprocal molar magnetic susceptibility $(\chi\)$_M$-\(\chi\)$_0)^{-1}(T)$ c) change of the Curie-Weiss temperature $\theta_{CW}$ and d) effective magnetic moment $\mu_{eff}$ with respect to composition ($x$) for Nd$_{1-x}$La$_x$NiC$_2$ solid solution. The inset to panel a): expanded view of the low temperature region for selected $x$ = 0.2 and $x$ = 0.6 compounds. Arrows indicate the transition temperatures T$_N$ and $T_N^*$. Yellow solid lines in panels a) and b) represent Curie-Weiss fit for NdNiC$_2$ (see text for details). Dashed lines in c) and d) are guides to the eyes, while black arrow indicates the breakdown of linearity in $\mu_{eff}(x)$.  }
  \end{figure*}
\indent
The temperature dependence of the dc molar magnetic susceptibility $\chi_M$ for the whole Nd$_{1-x}$La$_x$NiC$_2$ series was measured in the temperature range 1.9 - 300 K with $\mu_0H$ = 1 T applied magnetic field. Results for representative samples with $x\leq$ 0.9 are presented in Fig. \ref{AFM}a) whereas Fig.  \ref{AFM}b) depicts reciprocal molar susceptibility as a function of temperature. 
At high temperatures all Nd$_{1-x}$La$_x$NiC$_2$ compounds show paramagnetic behavior. Upon cooling, at low temperatures one can observe a sharp maximum associated with an antiferromagnetic transition (for $x$ ranging from $x$ = 0 to $x$ = 0.5). The Néel temperature, initially $T_N$ = 17 K for NdNiC$_2$ ($x$ = 0)(in agreement with ref. \cite{Roman2018_2}), decreases with the rise in La concentration ($x$) and for $x$ = 0.5 reaches $T_N$ = 9.5 K. Starting from $x\geq$ 0.6 the shape of the anomaly begins to broaden and finally for $x\geq$ 0.7 the transition is no longer observed in dc mode at applied field of $\mu_0H$ = 1 T. To distinguish between these two types of magnetic crossover, the Néel temperature is marked as $T_N$ for the $x$ range with a sharp character of transition and $T_N^*$ for the region where the accompanying features are more blurred. $T_N$ and $T_N^*$ were estimated as the maximum of the temperature derivative of the real part of magnetic susceptibility multiplied by the temperature $\frac{d(\chi_M'T)}{dT}$ (not shown here). To depict the contrast between these behaviors in Fig. \ref{AFM}a) (inset) shows the plots for $\chi_M(T)$ for $x$ = 0.2 and $x$ = 0.6, representative for sharp and blurred transition regions, respectively. The difference between them is likely associated with a weakening of the AFM interactions and an increase of spin disorder. This behavior differs from the results obtained for Sm$_{1-x}$La$_x$NiC$_2$ \cite{Prathiba2016}, and SmNiC$_{2-x}$B$_x$ \cite{Morales2014}, where weak doping initially causes a slight increase of Curie temperature $T_C$ followed by more abrupt suppression of FM for higher doping rates. 

In Nd$_{1-x}$La$_x$NiC$_2$, above the AFM transition temperature, all $(\chi\)$_M$-\(\chi\)$_0)^{-1}$ plots show an approximate linear dependence with T, indicating the relevance of the Curie-Weiss law expressed by the following equation:
\begin{equation}
\label{CurieWeiss}
\chi (T) = \frac{C}{T-\theta_{CW}} + \chi_0
\end{equation}
where C is the Curie constant, \(\theta\)$_{CW}$ is the Curie-Weiss temperature and \(\chi\)$_0$ is the temperature independent magnetic susceptibility (in this case coming both from the sample and the sample holder). The Curie constant is related to the effective magnetic moment \(\mu\)$_{eff}$ as shown in eq. 2: 
\begin{equation}
\label{EffectiveMoment}
\mu_{eff} = \sqrt{\frac{3 C k_B}{{\mu_B}^2 N_A}}
\end{equation}
where k$_B$ is the Boltzmann constant, N$_A$ is the Avogadro number and \(\mu\)$_B$ is the Bohr magneton. 
The fit with eq. 1 allowed the determination of the Curie-Weiss temperature and Curie constant which was used to calculate the effective magnetic moment \(\mu\)$_{eff}$ (an exemplary fit in the temperature range 50-300 K to the data for NdNiC$_2$ is shown with a solid yellow line in Fig. \ref{AFM}a) and b)). The Curie-Weiss temperature \(\theta\)$_{CW}$, along with the effective magnetic moment \(\mu\)$_{eff}$, are presented in Fig. \ref{AFM} c) and d), respectively.
\\
\indent
Upon the consequent increase of the La content in Nd$_{1-x}$La$_x$NiC$_2$ solid solution, the \(\theta\)$_{CW}$ starts to lower its absolute value from $\vert\theta_{CW}\vert$ = 22.9 K  for NdNiC$_2$ \cite{Roman2018_2} reaching almost zero value for $x$ = 0.4 which indicates a weakening of the AFM interactions between spins. This seems to be consistent with the decreasing concentration of magnetic Nd ions. For $x$ = 0.1 one can notice the deviation from the general trend for \(\theta\)$_{CW}$, however the origin of this anomaly is not clear. By further replacing Nd by La ions one should expect a continuous weakening of the magnetic interactions, while the Curie-Weiss temperature unexpectedly turns to more negative values up to \(\theta\)$_{CW}$ = -29.6 K for Nd$_{0.1}$La$_{0.9}$NiC$_2$.  For $x>$ 0.9 the absolute value of the Curie-Weiss temperature begins to diminish with a quasi-linear manner which coincides with the appearance of the superconducting state in compounds with high La content range ($x\geq$ 0.96). The gradual dilution of the Nd ions network with non-magnetic La alone is not sufficient to explain either the Curie-Weiss temperature approaching 0 for $x$ = 0.4, where the magnetic order still persists, or the sudden return of $\theta_{CW}$ to more negative values as the La content is further increased (0.4 $<x<$ 0.9). The presence of such extremum points to an increase in the role of magnetic fluctuations or a more complex evolution of interactions between local magnetic moments.
\\
\indent
The effective magnetic moment \(\mu\)$_{eff}$ varies with $x$ in an approximatly linear manner up to $x$ = 0.9 (see Fig. \ref{AFM}d)). The value of \(\mu\)$_{eff}$ decreases from 4.1\(\mu\)$_B$ \cite{Roman2018_2} for NdNiC$_2$ with increasing La concentration ($x$) in Nd$_{1-x}$La$_x$NiC$_2$, which is consistent with the $f$ electron number reduction caused by La substitution in place of Nd atoms. For $x>$ 0.9 (marked by arrow) \(\mu\)$_{eff}$ ceases to change linearly and drops abruptly towards a zero value for nonmagnetic LaNiC$_2$. This rapid fall of \(\mu\)$_{eff}$ is concomitant with the return of \(\theta\)$_{CW}$ towards zero. 

\begin{figure}[ht]
\includegraphics[angle=0,width=1.0\columnwidth]{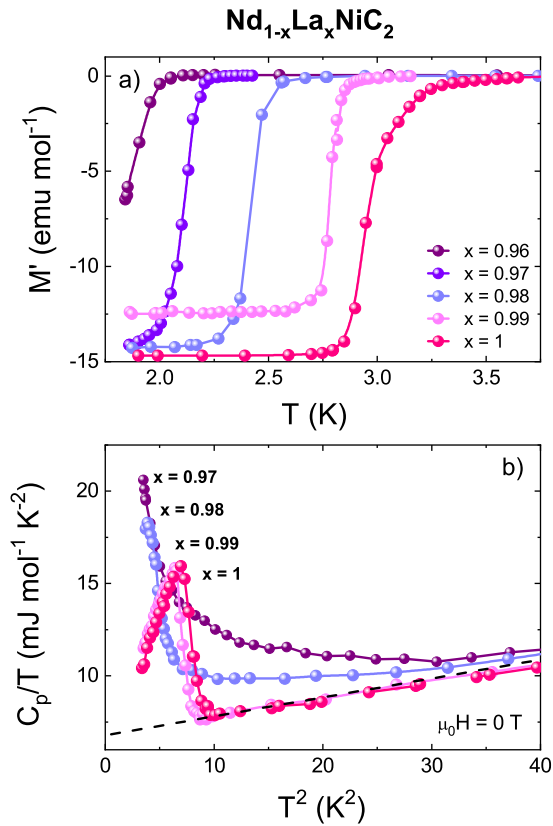}
 \caption{\label{SC} a) The real part of the molar magnetization $M'(T)$ and b) heat capacity over temperature C$_p$/T (T$^2$) for the superconducting samples ($x\geq$ 0.96) from Nd$_{1-x}$La$_x$NiC$_2$ series. A black dashed  line represents a fit to C$_p$/T = \(\gamma\)+{\(\beta\)}T$^2$ in the normal state of the low temperature region for the $x$ = 0.99 sample.}
  \end{figure}

Superconductivity appears beyond the point at which the AFM is completely suppressed. The superconducting transition is revealed by the temperature dependence of the real part of the ac molar magnetization $M'(T)$ as depicted in Fig. \ref{SC}a). 
A sharp diamagnetic drop in the magnetization is observed for La rich compounds and the critical temperature increases with $x$ from $T_{SC}$ = 1.98 K for Nd$_{0.04}$La$_{0.96}$NiC$_2$ to $T_{SC}$ = 3 K for LaNiC$_2$. Note that superconductivity persists only for small amounts of magnetic Nd dopant, which act as strong Cooper pairs breaking centers.
\\
\indent
 \begin{figure*}[ht]
\includegraphics[angle=0,width=2.1\columnwidth]{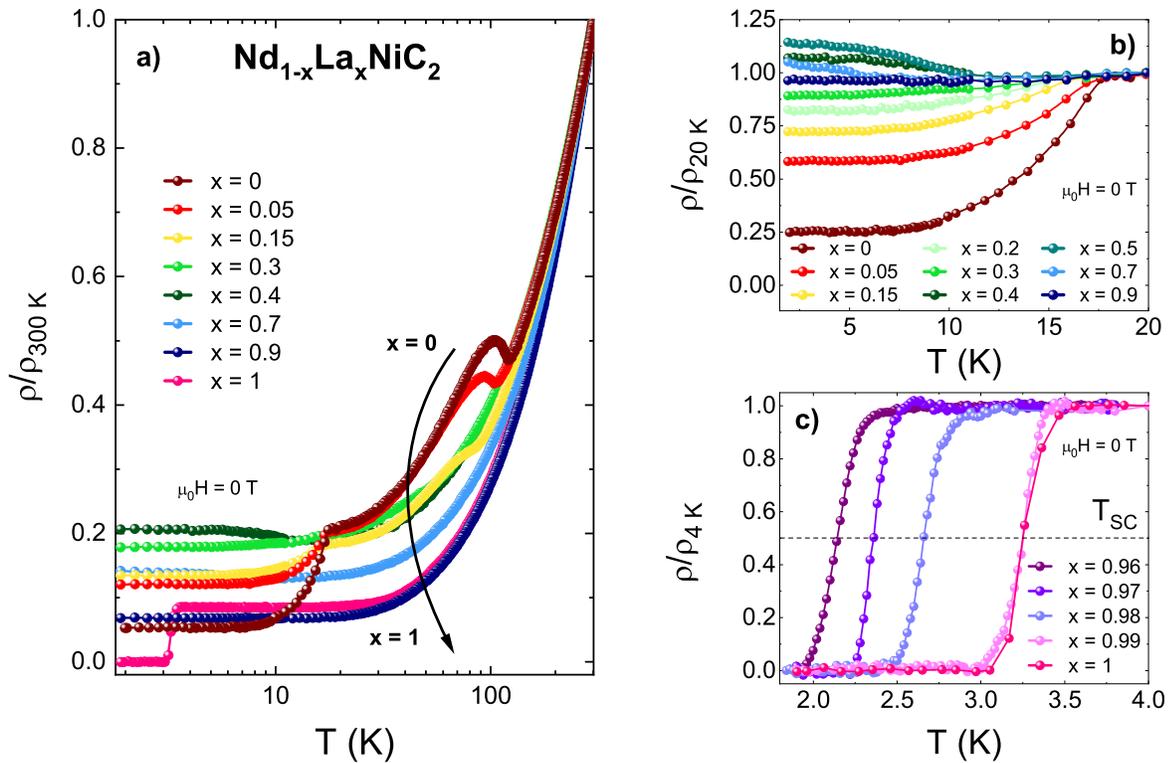}
 \caption{\label{resist} a) The thermal dependence of the normalized electrical resistivity for selected compounds from Nd$_{1-x}$La$_x$NiC$_2$ solid solution in the temperature range from 1.9 to 300 K. For clarity, the scale for horizontal axis is logarithmic. Panels b) and c) show the expanded view of the low temperature region for $x\leq$ 0.9 and $x>$ 0.9, respectively.}
  \end{figure*}
In order to confirm the volume character of the superconducting transition, specific heat capacity measurements were performed, and $\frac{C_p}{T}(T^2)$ is presented in Fig. \ref{SC}b). For $x$ = 1 a sharp superconducting transition is visible at $T_{SC}$ = 3 K and as Nd ions are introduced into LaNiC$_2$, the critical temperature decreases with simultaneous enhancement of lambda-shape specific heat jump at the transition.  Finally, for $x\leq$  0.97, although $\frac{C_P}{T}$ abruptly increases at low temperature, no maximum is observed above 1.9 K. This feature is \textit{a priori} unexpected, since one rather expects the weakening of the superconducting transition as $T_{sc}$ is depressed and thus suggests the occurrence of additional mechanism contributing to low temperature specific heat. 
The experimental data points of the normal state were fitted using the formula:
\begin{equation}
\label{Cp_eq}
\frac{C_p}{T} = \gamma+\beta T^2
\end{equation}
where the first and second terms in the right side of eq. \ref{Cp_eq} represent the electronic and lattice contribution to the specific heat, respectively.
It is worth noting that the curves for 0.97 $\leq x \leq$ 0.99 present a similar slope and coincide with each other above $T \simeq$ 7.5 K, indicating a barely noticeable change in thermodynamic parameters above the superconducting transition.  The fit for $x$ = 0.99 (black dashed line in Fig. \ref{SC}b)) provides values of the Sommerfeld coefficient \(\gamma\) = 6.8(1) mJ mol$^{-1}$ K$^{-2}$ and \(\beta\) = 0.102 mJ mol$^{-1}$ K$^{-4}$. 
The Debye temperature \(\theta\)$_D$ was estimated using a simple Debye model for the lattice contribution:
 \begin{equation}
\label{thetaD_eq}
\theta_{D} = {\left(\frac{12 \pi^4}{5\beta}nR\right)}^{\frac{1}{3}}
\end{equation}
where R = 8.314 mol$^{-1}$ K$^{-1}$ and $n$ is the number of atoms per formula unit (here $n$ = 4). The calculated  \(\theta\)$_D$ shows a relatively high value of 423 K due to the presence of  light carbon atoms. The obtained Sommerfeld coefficient and the Debye temperature are close to the values determined for LaNiC$_2$ (fit not shown) which are \(\gamma\) = 6.6(0) mJ mol$^{-1}$ K$^{-2}$ and \(\theta\)$_D$ = 427 K, respectively, also in agreement with previous reports \cite{Prathiba2016}.
\\
\indent
The results of electronic transport measurements for the whole Nd$_{1-x}$La$_x$NiC$_2$ ($0\leq x\leq 1$) series are presented in Fig. \ref{resist}a), where resistivity values are normalized to those at 300 K for comparison. Panels b) and c) delineate the resitivity curves \(\rho\)/\(\rho\)$_{20 K}(T)$ for compounds showing AFM and \(\rho\)/\(\rho\)$_{4 K}(T)$ for samples exhibiting superconductivity, respectively. 
\\
\indent
The character of the resistivity evolves with $x$. At high temperatures, all compounds show typical metallic character with $\frac{d\rho}{dt}<0$. For $x<0.4$, CDW metal-metal transition is observed at Peierls temperature with a maximum value of $T_{CDW}$ = 121 for NdNiC$_2$ and gradually lowering as Nd ions are replaced by La.  The magnitude of the resistivity maximum accompanying the CDW transition decreases together with the Peierls temperature. In Nd$_{0.7}$La$_{0.3}$NiC$_2$ this anomaly is visible only as a weak inflection of the resistivity curve at $T_{CDW}$ = 53.5 K, while for higher doping rates ($x>$ 0.3) the Peierls transition is no longer observed and the metallic character of the conductivity is preserved down to $T_N$ or respectively $T_{SC}$.
At N\'{e}el temperature a rapid drop of resistivity is observed for compounds with $x\leq$ 0.3, thus those exhibiting a CDW. For compounds with 0.4 $\leq x \leq$ 0.8, the resistivity starts growing as the temperature is decreased below $T_N^*$. For $x$ = 0.9 small increase of $\rho(T)$ is observed at low temperatures, however the magnetic susceptibility measurements do not detect any signatures of magnetic transition above $T$ = 1.9 K. (see Fig. \ref{resist}b) for the expanded view of low temperature resistivity curves). For $x>$ 0.9, where the antiferromagnetic ground state is suppressed, the low temperature behavior of resistivity evolves again, and once more shows a decrease, this time reaching the zero value due to the superconducting transition (see Fig. \ref{resist}c)). Such a sharp crossover is visible for compounds with La content $x\geq$ 0.96 with critical temperatures ranging from $T_{SC}$ = 2 K for $x$ = 0.96 to $T_{SC}$ = 3.2 K for $x$ = 1, thus slightly higher than estimated from magnetic and heat capacity measurements. For $x\leq$ 0.97, despite a pronounced increase of $\frac{C_p}{T}$ at lowest temperatures, no clear maximum can be observed above 1.9 K.

 \begin{figure*}[ht]
\includegraphics[angle=0,width=2.0\columnwidth]{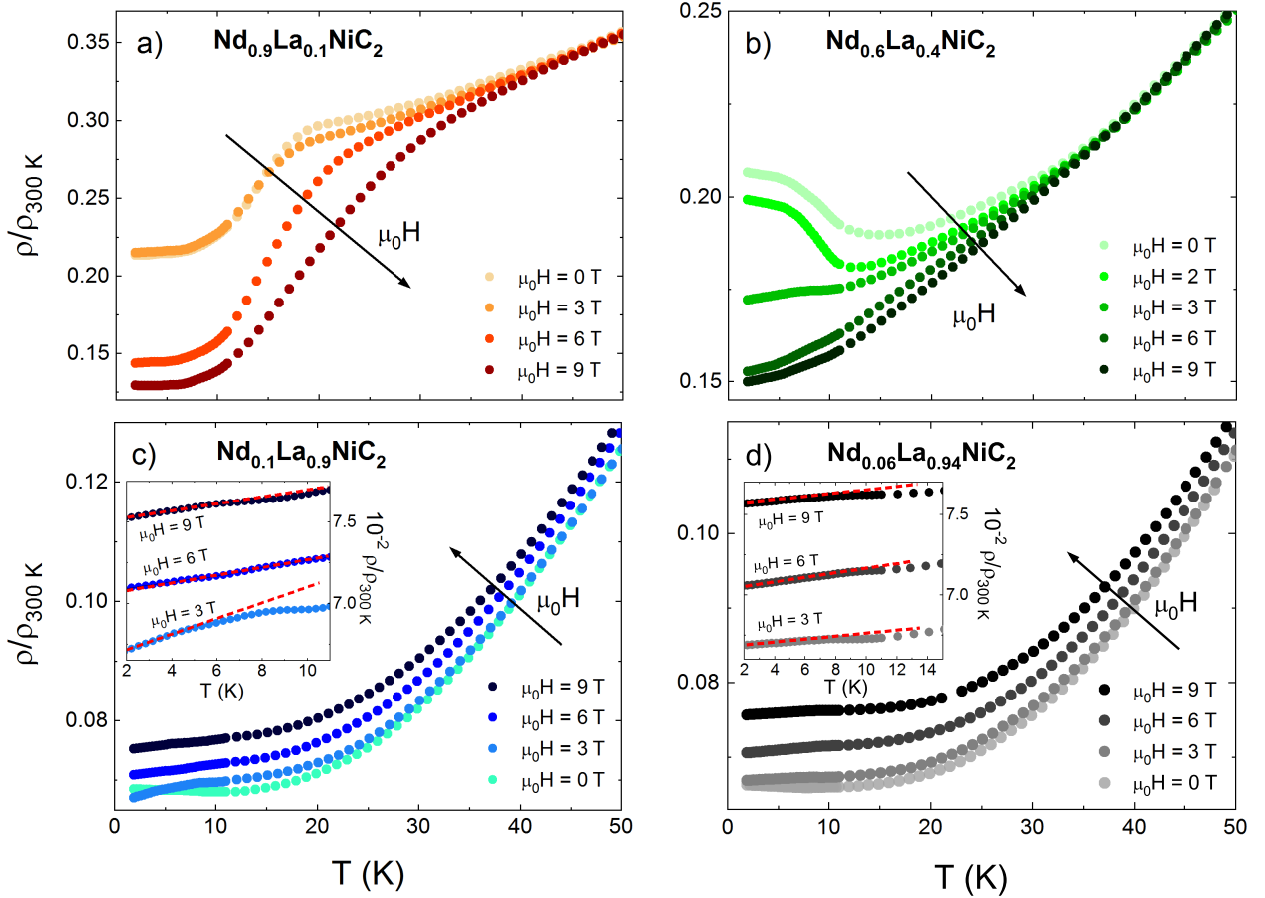}
 \caption{\label{MR} The thermal variation of the normalized electrical resistivity measured at various magnetic fields for selected Nd$_{1-x}$La$_x$NiC$_2$ samples a) $x$ = 0.1, b) $x$ = 0.4, c) $x$ = 0.9, and d) $x$ = 0.94. Insets to panels c) and d) show the expanded view of the region with linear $\rho(T)$. The dashed lines are guide for the eye.}
  \end{figure*}

The increase in the resistivity below the N\'{e}el temperature stands in contrast with the behavior seen in the Sm$_{1-x}$La$_x$NiC$_2$ solid solution\cite{Prathiba2016}, where the drop of resistivity was observed at Curie temperature even for the intermediate compounds where the charge density wave was already suppressed. Previously, for parent NdNiC$_2$ the decrease of the resistivity at the magnetic ordering temperature was attributed both to the partial suppression of the charge density wave, concomitant with the release of condensed carriers, and the reduction of the spin disorder together with the underlying scattering rate \cite{Yamamoto2013, Kolincio2017, Lei2017}. This is also true for GdNiC$_2$ \cite{Shimomura2016, Kolincio20161, Hanasaki2017} and their solid solution Nd$_{1-x}$Gd$_x$NiC$_2$ in the whole $x$ range\cite{Roman2018_2}. A stronger effect was observed in SmNiC$_2$, where the charge density wave was completely suppressed\cite{Shimomura2009, Hanasaki2012, Lei2017, Kim2012}. In Nd$_{1-x}$La$_x$NiC$_2$ the resistivity drop below $T_N$ is observed only for $\leq$ 0.3, where the emergence of the CDW was detected, it is then reasonable to assume that this effect is, at least partially, caused by the  weakening or the destruction of the charge density wave. Nevertheless, one should not underestimate the role played by the resistivity component associated with the spin disorder scattering. The reduction in the resistivity at $T_N$ has also been observed in isostructural CeNiC$_2$\cite{Kolincio2017}, deprived of the Peierls transition, which reflects the impact of spin fluctuations on the resistivity in the vicinity of $T_N$. Although for low values of $x$ both terms appear to be relevant for high Nd concentrations, in the absence of a CDW for $x\geq$ 0.4, the spin disorder is expected to play a decisive role in determining the form of $\rho(T)$ beneath the N\'{e}el temperature. It is surprising, however, not to observe the resistivity lowering upon entering the magnetically ordered state, which is expected to be concomitant with reduction of spin disorder as in CeNiC$_2$. The adverse direction of the resistivity evolution below $T_N$ suggests rather the enhancement of the spin fluctuations instead of their condensation to long range antiferromagnetism. Next to the spin disorder, the increase of resistance in this temperature range can partially originate from the Kondo effect with dispersed magnetic ions acting as scattering centers\cite{Kondo1964, Jones2007}. We do not however find a logarytmic dependence of $\rho$ as $T\rightarrow0$, which is a characteristic feature of Kondo scattering with magnetic impurities \cite{Kobayashi2010, Rotella2015}. The growth of $\rho(T)$ below $T_N$ observed in Nd$_{1-x}$La$_x$NiC$_2$ does not lead to a maximum as reported in antiferromagnets with dominance of Kondo interactions \cite{Nakashima_2017, Nakamura_2015}. In these systems, resistivity drops significantly below the magnetic ordering temperature due to the suppression of spin disorder scattering as in regular AFM metals. The absence of such a drop and continuous increase of $\rho(T)$ as $T \rightarrow 0$ suggests that the spin disorder scattering is a dominant mechanism, despite the increase of the role played by Kondo coupling in the terms of magnetic properties. 
  The alternative scenario, the superzone boundary effect due to the mismatch between magnetic and crystalographical Brillouin zones observed in some AFM systems \cite{Klimczuk_2015, Elliott_1963} appears not to be relevant, since the resistivity upturn is not seen for Nd$_{1-x}$La$_x$NiC$_2$ with high Nd concentrations and the Brillouin zone is not expected to significantly evolve between NdNiC$_2$ and LaNiC$_2$ since there is no drastic changes to the lattice parameters (see figure \ref{XRD}).
\\
  \begin{figure}[ht]
\includegraphics[angle=0,width=1.0\columnwidth]{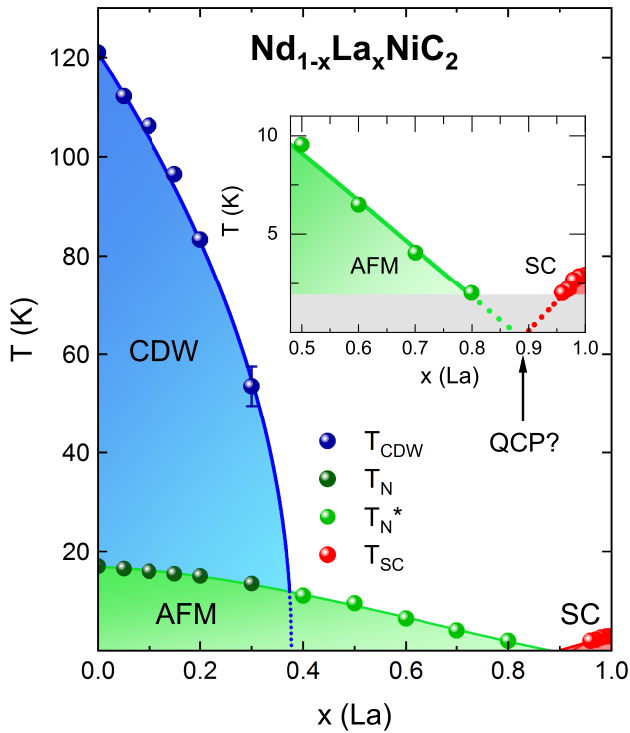}
 \caption{\label{Diagram}  Phase diagram of temperature vs. composition ($T - x$) for Nd$_{1-x}$La$_x$NiC$_2$ series.  CDW phase with Peierls temperature $T_{CDW}$ is represented by gradient blue color. AFM phase with $T_N$ and $T_N^*$ by gradient green color and SC with $T_{SC}$ phase by red color. The N\'{e}el temperature for the AFM crossovers of distinct characters is marked by dark green points for the region in which the transition is sharp ($T_N$) and light green color, where the AFM transition is blurred and accompanying anomalies are weakened ($T_N^*$). Blue line is the fit to $T_{CDW}(x)$ with equation \ref{BCS}. Inset: expanded view for the low temperature region to highlight the collision of AFM and SC regions at a possible quantum critical point marked by red arrow.}
  \end{figure}
\indent
Complementary information on spin disorder can be obtained from magnetoresistance (MR) measurements. In Fig. \ref{MR} we compare the influence of magnetic field on transport properties of selected members of the Nd$_{1-x}$La$_x$NiC$_2$ family, representative for the regions with distinct low temperature resistivity behaviors. At temperatures far above the magnetic ordering, the magnetic field has a negligible impact on the resistivity. A stronger effect is visible as $T$ is lowered. For $x$ = 0.1 (Fig. \ref{MR}a)) the negative magnetoresistance term prevails both above and below N\'{e}el temperature. The character of the MR in this compound is reminiscent with the features seen in the parent NdNiC$_2$, where the suppression of the charge density wave plays a crucial role in the magnetoresistive effects \cite{Yamamoto2013, Kolincio2017, Lei2017}. By this analogy, it is reasonable to assume that the destruction of CDW is responsible for at least a part of MR in Nd$_{0.9}$La$_{0.1}$NiC$_2$. It is then not straightforward to isolate the spin scattering term from the whole magnetoresistance picture. 
For $x$ = 0.4 (Fig. \ref{MR}b)) however, the CDW transition is no longer observed, thus the spin fluctuations are expected to be the main driving force of the magnetoresistance\cite{Mazumdar1997, Yamada1972, Akhavan1976}. The application of a magnetic field reduces the height of the resistivity hump observed below $T_N$, which can be attributed to a partial reorientation of the magnetic moments and a reduction of the magnetic entropy. The magnitude of this effect grows as the magnetic field is increased and at $\mu_0H$ = 3 T the resistivity maximum is completely suppressed. Application of a stronger magnetic field continues to suppress the spin disorder and  drives the resistivity even lower. Eventually, at high $\mu_0 H$, $\rho$ ceases to decrease upon further increasing the magnetic field, presumably due to a final quench of the spin fluctuations by the field induced ferromagnetic crossover.
For $x$ = 0.9 (Fig. \ref{MR}c)), showing no magnetic ordering down to 1.9 K the application of a magnetic field suppresses the weak upturn of zero field resistivity curve as $T\rightarrow 0 $, unveiling a remarkably linear $\rho(T)$ dependence. Further increase of $\mu_0H$ beyond this point increases the value of resistivity, presumably due to the ordinary Lorentz mechanism, yet the linear $\rho(T)$ behavior is conserved at higher fields. The expanded view for the region with $\rho \sim T$ is highlighted in the inset of figure \ref{MR}c). For higher La concentrations, this term is less pronounced, as seen for in Fig. \ref{MR}c). Finally for $x\geq$ 0.97 the linearity is no longer observed within the experimental resolution.

The transport results corroborate the magnetization measurements, showing a gradual softening of the features associated with the AFM transition as the Nd content in Nd$_{1-x}$La$_x$NiC$_2$ is decreased (thus $x$ is increased). Both series of results reveal symptoms of disordered aniferromagnetic behavior for $x\geq$ 0.4.  Interestingly, this crossover coincides with the vanishing of the charge density wave state; compounds with a Peierls transition reveal more ordered character than those in which the CDW is absent. It is plausible then to attribute this effect to the recently suggested stabilization of antiferromagnetism by charge density wave via Fermi surface nesting enhancement of the RKKY interaction between magnetic ions and the formation of spin density wave in GdNiC$_2$, NdNiC$_2$ and their solid solutions\cite{Kolincio20161, Hanasaki2017, Roman2018_2}. When RKKY interaction is no longer enhanced by charge density wave, and its strength is weakened, the Doniach picture\cite{Doniach1977} predicts the increase of the role played by Kondo interaction as the RRKY mechanism is weakened. This scenario can also explain the complex character of the $\theta_{CW}(x)$ curve. The initial decrease of $|\theta_{CW}|$ for $x<$ 0.4 corresponds to the region, where CDW is gradually suppressed, which stands for the weakening of the RKKY mechanism and as charge density wave dissapears,  paramagnetic Curie-Weiss temperature approaches zero. The further increase of La content beyond this point results in the inflection of $|\theta_{CW}|(x)$ in the region where the antiferromagnetic state is still present, alhough thermal dependence of magnetic susceptiblity and electrical resistance reveal signatures of magnetic fluctuations and a certain degree of disorder corresponding to them. Such an increase of $|\theta_{CW}|$ is expected to reflect the growth of the Kondo energy  \cite{Krishna-murthy1975, Lai2018} that starts taking control over the magnetic ordering. The existence of magnetic fluctuations as well as the competition between Kondo and RKKY interactions can additionally lead to quantum critical behavior of magnetic ordering\cite{Knebel1996, Lai2018}.
\\
\indent
To summarize the results from both magnetic and transport measurements, they were used to construct the phase diagram for the Nd$_{1-x}$La$_x$NiC$_2$ ($0\leq x\leq 1$) series which is depicted in Fig. \ref{Diagram}. 
The blue color represents the region in which CDW is observed with the Peierls temperature gradually suppressed from $T_{CDW}$ = 121 K for NdNiC$_2$ with increasing of the La concentration. 
$T_{CDW}(x)$ is succesfuly described by mean-field power law function characterizing the influence of chemical pressure \cite{Jaramillo2009, Montverde2013, Morales2014}:
\begin{equation}
\label{BCS}
T_{CDW} = T_{CDW}(0)\sqrt{1-\frac{x}{{x_c}}}
\end{equation}
where $T_{CDW}$(0) is the temperature of CDW transition for $x$ = 0, $x_c$ is La content corresponding to $T_{CDW}$ = 0 K.  Constraining the fit with constant $T_{CDW}$ = 121 K for undoped NdNiC$_2$ gives the value of $x_c$ = 0.38, slightly below the first point ($x$ = 0.4) at which the CDW transition is no longer observed. The fit with equation \ref{BCS} is shown in figure \ref{Diagram} as a blue line. The AFM region is represented by green color, dark and light green points stand for $T_N$ and $T_N^*$ respectively. The decrease of N\'eel temperature is not as steep as in the case of $T_{CDW}$ and for $x$ = 0.9 AFM is no longer observed above $T$ = 1.9 K. A further increase of La concentration results in an almost immediate onset of superconductivity (represented by the color red in fig. \ref{Diagram}), which rises for $x\geq$ 0.96  and critical temperature starts to increase with $x$. The inset of fig. \ref{Diagram} presents an expanded view of the antiferromagnetic and superconducting region. Curves describing the $x$ dependence of N\'eel and critical temperatures can be extrapolated to $T$ = 0 K. Interestingly both lines intersect at zero temperature near $x^*$ = 0.88, suggesting the putative existence of the AFM quantum critical point in Nd$_{1-x}$La$_x$NiC$_2$ series. Typically, the quantum criticality is accompanied by characteristic features in electrical resistivity in the vicinity of QCP\cite{Sachdev2011, Coleman2005}. This effect is expected to be pronounced by the softening of the temperature dependence of resistivity via reduction of exponent $p$ in formula \ref{res_fit}:
\begin{equation}
\label{res_fit}
\rho(T) = \rho_0+AT^p
\end{equation}
where $\rho_0$ is residual resistivity and the second term stands for the resistivity component dependent on temperature with $p$, indicating the prevailing type of scattering - $p$ = 1 is expected in quantum critical regime \cite{Gooch2009, Stewart2001}.

The lineartity of $\rho(T)$ was reported in a wide concentration range near ferromagnetic QCP in SmNiC$_{1-x}$B$_x$\cite{Morales2014} and SmNiC$_2$ under pressure\cite{Woo2013}. The signatures of such an effect in Nd$_{1-x}$La$_x$NiC$_2$ are seen only in $\rho(T)$ curves measured in the presence of external magnetic field, for 0.6 $<x<$ 0.97, close to presumed QCP at $x$ = 0.88. A plausible scenario is, that the linear, non-Fermi liquid behavior buried beneath the low temperature upturn of resistivity is uncovered by the magnetic field quenching the spin disorder scattering. It shall, however, be mentioned that typically the critical region with a non-Fermi liquid behavior is confined to a narrow vicinity of QCP \cite{Narayan2019, Grigera2001}, while the linearity in transport properties of Nd$_{1-x}$La$_x$NiC$_2$ is seen in a asymmetric zone, extended in the direction of low La concentrations.
Although there are exceptions, as a rather wide quantum critical region accompanying the transition from spiral to ferromagnetic phases in ZnCr$_2$Se$_4$ \cite{Gu2018}, in Nd$_{1-x}$La$_x$NiC$_2$ the linear $\rho(T)$ dependence near the AFM - SC crossover can also originate from other factors - such as a direct impact of spin scattering. Therefore, this effect cannot be treated as a clear evidence of QCP, but rather as a clue pointing towards its possible occurrence.

The possibility of quantum critical behavior even in complete absence of resistivity softening has been recently concluded based on the clear increase in $\frac{C_p}{T}$ as $T \rightarrow 0$, on the superconducting side of presumed QCP in Sm$_{1-x}$La$_x$NiC$_2$\cite{Prathiba2016}. 
An analogous situation is observed in Nd$_{1-x}$La$_x$NiC$_2$, where the magnitude of the specific heat jump near the onset of superconductivity notably increases despite the critical temperature being gradually suppressed as La atoms are substituted with Nd. Since the SC is weakened, the enhancement of $\frac{C_p}{T}$ likely stems from fluctuations 
emerging at low temperature. On the one hand, their origin can be purely magnetic, due to the vicinity of the AFM state. On the other hand, such a singular amplification of specific heat in the low temperature limit when $x \rightarrow x^*$ is a typical feature for critical order parameter fluctuations in the vicinity of QCP \cite{Zhu2003, Steppke2013, Westerkamp2009, Sheppard2011}. It is possible then, to attribute the growth of low temperature $C_p$, at least partially to the latter term.
   To unambiguously clarify the nature of the AFM to SC crossover, the transport and specific heat measurements must be extended to He$^3$ temperatures. Alternative methods are nuclear magnetic resonance (NMR) \cite{Kinross2014}, neutron diffraction or muon spectroscopy \cite{Lee2017} allowing to directly confirm (or deny) the quantum criticality near the point of contact of these two types of order parameters.
\section{Conclusions}
\indent
The Nd$_{1-x}$La$_x$NiC$_2$ ($0\leq x\leq 1$) solid solutions have been synthesized. By consequent replacement of Nd with La ions, the evolution from NdNiC$_2$ revealing both the CDW and AFM state to noncentrosymmetric unconventional superconductor LaNiC$_2$ has been investigated. The structural changes caused by doping-induced chemical pressure are manifested in linear variation of structural parameters in agreement with Vegard's law. 
The substitution of La in Nd positions results in an abrupt suppression of charge density wave and for La content higher than $x$ = 0.4 this ordering is no longer observed.  
We have found that as long as the CDW state is preserved, the AFM ground state shows strong anomalies in magnetic susceptibility and transport properties. With the further increase of La concentration, for compounds where the CDW is completely suppressed, the features associated with AFM transition become smeared, which is accompanied with the signatures of spin disorder leading to resistivity rise beneath the temperature of magnetic anomaly and negative magnetoresistance. Such crossover suggests a strong role played by charge density wave in the stabilization of antiferromagnetism, via formation of spin density wave in the presence of strong local magnetic moments. The gradually suppressed magnetism is replaced by superconductivity observed for La-rich compounds (for $x>$ 0.96), where the critical temperature quickly diminishes with a small amount of the magnetic Nd ions. The results of magnetic and transport properties of Nd$_{1-x}$La$_x$NiC$_2$ ($0\leq x\leq 1$) series are summarized in a comprehensive $T - x$ phase diagram. The extrapolation of curves following the variations of characteristic temperatures for antifferomagnetic order ($T_N$) and superconductivity ($T_{sc}$) suggests the putative existence of a critical point near $x^*$ = 0.88 where these two entities subside to zero temperature. The characteristic features that can be seen as signatures of quantum criticality can be found in specific heat and transport properties.
\\
\section{Acknowledgments}
 The authors gratefully acknowledge the financial support from National Science Centre (Poland), grant number:  UMO-2015/19/B/ST3/03127. Authors would also like to thank to H. Walker (ISIS) and N. Runyon for their helpful advice.

%

\end{document}